\documentstyle[aps,prl,epsfig,special]{revtex}

\newcommand{\ket}[1]{\left|#1\right\rangle}

\begin{document}
\draft
\title{A Model for Ferromagnetic Nanograins with Discrete
  Electronic States}
\author{Silvia Kleff${}^{1}$, Jan von Delft${}^{1,2}$, 
Mandar M. Deshmukh${}^{3}$, and D. C. Ralph${}^{3}$}
\address{${}^1$Institut f\"ur Theoretische Festk\"orperphysik,
Universit\"at Karlsruhe, 76128 Karlsruhe, Germany\\
${}^2$Physikalisches Institut, Universit\"at Bonn, 53115 Bonn,
Germany\\
${}^3$Laboratory of Atomic and Solid State Physics,
Cornell University, Ithaca, NY, 14853}
\date{(Submitted: April 2, 2001)}
\maketitle
\begin{abstract}
  We propose a simple phenomenological 
 model for an ultrasmall ferromagnetic grain,
  formulated in terms of the grain's discrete energy levels.  We compare the
  model's predictions with recent measurements of the discrete tunneling
  spectrum through such a grain. The model can qualitatively account for 
the observed features if we assume (i) that the anisotropy energy 
varies among different eigenstates of 
one grain, and (ii) that nonequilibrium
spin accumulation occurs. 
\end{abstract}
\narrowtext

What are the properties of individual quantum states in the electronic
excitation spectrum of a nanometer-scale ferromagnetic particle?  This
is becoming an increasingly important question, since the size of
memory elements in magnetic storage technologies is decreasing
extremely rapidly \cite{johnson}, and particles as small as 4 nm are
coming under investigation \cite{murray}. In this size regime, the
excitation spectrum becomes \emph{discrete}; indeed, Gu\' eron,
Deshmukh, Myers and Ralph (GDMR) \cite{gueron} have recently succeeded
to resolve individual quantum states in the spectrum of ferromagnetic
Cobalt nanograins, using single-electron tunneling spectroscopy. They
found \emph{complex nonmonotonic and hysteretic energy level shifts}
in an applied magnetic field and an unexpected \emph{abundance of
  low-energy excitations}, which could not be fully understood within
the simple models used for ferromagnetic nanograins so far
\cite{gueron,canali}.

In this Letter, we propose a phenomenological model for ferromagnetic
nanograins 
that is 
explicitly formulated in terms of the discrete
states occupied by the itinerant conduction electrons and capable of
qualitatively explaining the observed features. The model is similar
in spirit to that advanced independently by Canali and MacDonald
\cite{canali}, but our analysis includes two further ingredients 
beyond theirs: (i) mesoscopic fluctuations of the anisotropy
energy (i.e. it may vary among different eigenstates), 
and (ii) nonequilibrium spin accumulation.

{\em Experimental Results.}--- GDMR studied Co-particles 1-4 nm in
diameter.  Assuming a hemispherical shape, the number of atoms in such
grains is in the range $N_{\rm a} \approx 20$--1500, and the total
spin, $s_0 \approx 0.83 N_{\rm a}$ \cite{Papa}, thus is $ s_0 \approx
17-1250$.  In GDMR's devices, a grain is connected to two aluminum
electrodes via aluminum oxide barriers. Its tunneling conductance
consists of a series of distinct peaks (see Fig.~2 in \cite{gueron}),
whose positions yield a set of tunneling energies of the form
\cite{jvdralph} $\Delta E^{\pm}_{fi} \equiv E_f^{N\pm 1} - E_i^N \,$
each corresponding to the energy cost of some rate-limiting electron
tunneling process $|i\rangle^{N} \to |f\rangle^{N \pm 1}$ onto or off
the grain.  Here $|i\rangle^{N}$ denotes a discrete eigenstate, with
eigenenergy $E^{N}_i$, of a grain with $N$ electrons, etc.

As the magnetic field is swept, the resonances for Co-grains undergo
energy shifts and crossings (Fig.~3 in \cite{gueron},\cite{mandar}).
The resulting tunneling spectra have several properties that differ
strikingly from those of previously-studied
nonmagnetic Al and Au grains \cite{RBT,drago}:\\
(P1): Many \emph{more low-energy excitations} are observed than
expected: For all values of magnetic field, the mean level spacing is
$\bar d_{\rm obs} \lesssim 0.2$~meV.  This is much \emph{smaller} than
expected from the naive single-particle estimates $\bar d_{\rm maj}
\approx 4.6~{\rm eV}/s_0$ or $\bar d_{\rm min} \approx 1.2~{\rm
  eV}/s_0$ (with $s_0 \lesssim 1250$) for the majority- and
minority-spin mean level spacings near the Fermi energy of Co
\cite{bulkN,Papa}.
\\
(P2): In the small-field regime ($\mu_{0}H<0.2$ T), discontinuous
hysteretic switching occurs at a certain switching field $H_{\rm sw}$,
due to a sudden change in direction (henceforth called ``reversal'')
of the magnetic moment. Moreover, the $H$-dependence of tunneling
resonance energies has continuous non-monotonic variations, which
differ seemingly randomly from level to level (Fig.3 in
\cite{gueron},\cite{mandar}).
\\
(P3): In the large-field regime ($|H| \gg |H_{\rm sw}| $), the
resonance energies depend roughly \emph{linearly} on $H$, with
$H$-slopes that almost all have the \emph{same sign} for a given
grain; in particular, slopes of opposite signs due to Zeeman-splitting
of spin-up and spin-down levels \cite{RBT,drago} are not observed
(Fig.~4 in \cite{gueron},\cite{mandar}).

Point (P2) indicates immediately that an independent-electron approach
to the energy levels is not sufficient, because the energy of a given
state depends on the orientation of the magnetic moment produced by
all the electrons within the particle.  We shall argue that points
(P1) and (P3), too, are related to the many-electron spin structure
within the particle.

{\em Model Hamiltonian.}---We propose to model a nanoscale magnet with discrete
excitations by the following ``minimal'' Hamiltonian: 
${\cal H}={\cal H}_{0} + {\cal H}_{\rm C}+
{\cal H}_{\rm exch} +{\cal H}_{\rm Zee} + {\cal H}_{\rm anis} $,
where ${\cal H}_{\rm C}$ is the Coulomb charging energy
for a nanoparticle containing $N$ electrons, and
\begin{eqnarray}
\label{H0}
{\cal H}_0 & = & \sum_{j \sigma} \varepsilon_{j}
      c_{j \sigma}^\dagger c_{j \sigma} \; ,
\quad  {\cal H}_{\rm exch} = - {U \over 2} \vec S \cdot \vec S
   \; ,
\\
\label{Hanis}
{\cal H}_{\rm Zee}  & = &
-h S^z \; , \quad \qquad
  {\cal H}_{\rm anis} =
  - \: \sum_{ab} \sum_{ij} S^a_i {\cal K}^{ab}_{ij} S^b_j \; ,
\end{eqnarray}
\noindent
with $ h=g_{\text{eff}}\mu_{\rm B}\mu_{0} H$.  Here ${\cal H}_0$
describes the kinetic energy of a single band of single-electron
states $|j, \sigma \rangle$, labeled by a discrete index $j$ and a
spin index $\sigma = \mbox{$(\uparrow, \downarrow)$}$, with the spin
quantization axis chosen in the $z$-direction.  The exchange, Zeeman,
and anisotropy terms, ${\cal H}_{\rm exch}$, ${\cal H}_{\rm Zee}$ and
${\cal H}_{\rm anis}$, are functions of the level-$j$ spin operators $
S_j^a = \frac{1}{2} \sum_{\sigma'\sigma} c^\dagger_{j \sigma}
\sigma^a_{\sigma \sigma'} c_{j \sigma'} $ (where ${\sigma}^a$ are
Pauli matrices, with $a = x,y,z)$, so that $\vec S = \sum_j \vec S_j$
is the total spin vector.  ${\cal H}_{\rm exch}$ is a rotationally
invariant term which models the effects of an exchange field and
forces the system to adopt a non-zero total ground state spin, say
$s_0$.  On account of this term, spins aligned parallel or
antiparallel to $\langle \vec S\rangle $ may be thought of as forming
``majority'' and ``minority'' bands, which effectively rotate rigidly
together with the magnetization direction.  We shall take the mean
level spacings $(\bar d_{\rm maj},\bar d_{\rm min}$) near the
respective Fermi energies, and the exchange-splitting of the Fermi
energies, $\Delta_{\rm F} \equiv \varepsilon_{\rm F,maj} -
\varepsilon_{\rm F,min}$ ($\approx 2$~eV for Co), as characteristic
parameters of the model. The magnitude of $U$ may then be estimated as
$U \approx \Delta_{\rm F}/s_0$, since stability of the ground state
spin $s_0$ implies \cite{canali} the relation $\Delta_{\rm F} = U (s_0
+ 1/2) + d_0$, where $d_0 (\sim 1/s_0)$ is a small, grain-dependent
energy satisfying $- (\bar d_{\rm maj} - U/2) < d_0 < \bar d_{\rm min}
- U/2$.  ${\cal H}_{\rm Zee} $ describes the spin Zeeman energy in an
external magnetic field $\vec{H} = H \hat z$.  Finally, ${\cal H}_{\rm
  anis} $ models the combined effects of crystalline, shape and
surface anisotropies, etc, in terms of a hermitian, traceless tensor
[$\sum_a {\cal K}^{aa}_{ij} = 0$], which describes the energy cost for
rotating the various spins $\vec S_j$. We split the tensor into an
``average" and a ``fluctuating" part by writing ${\cal K}^{ab}_{ij} =
K^{ab} + k^{ab}_{ij}$.  The $K^{ab}$ part dominates since all levels
contribute coherently, and, assuming $K^{ab} \propto 1/ {\rm Vol}$, it
makes an extensive ($\propto {\rm Vol}$) contribution to the total
energy.  The simplest non-trivial form that this term might take is a
uniaxial anisotropy
\begin{equation}
\label{eq:uniaxial}
{\cal H}_{\rm uni} = -k_N (\vec S \cdot \hat n)^2 /s_0 \, ,
\end{equation}
where $\hat n$ is the unit vector in the easy-axis direction (at,
say, an angle $\theta$ from $\hat z$)
and $k_N (>0)$ is a volume-independent constant.
The fluctuating term $k^{ab}_{ij}$  causes the total anisotropy
energy to depend on which single-particle levels are occupied.
It is a new ingredient relative to previous models of magnetic switching,
which required only a single anisotropy energy function for the whole
system, as is appropriate when only the ground state magnetic properties are
pertinent \cite{wernsdorfer}. 

{\em Basis states.}--- It is convenient to use the eigenstates of 
${\cal H} ({\cal K}=0)$ to construct a set of ``bare" basis states.  
Since $[{\cal H},\vec{S}]=0$, these states can be grouped into spin
multiplets that are labeled by their $\vec S \cdot \vec S$ and $S_z$
eigenvalues, say $s(s+1)$ and $m$. For example, the bare ground state
of ${\cal H}({\cal K}=0)$ for given $N$, $s$ and $h \,(>0)$, say
\begin{eqnarray}
\label{highestsstate}
\ket{s,s}^{N}_{0}
& \equiv & \prod_{j=1}^{n_{\uparrow}}c^{\dagger}_{
j\uparrow} \; \prod_{j=1}^{n_{\downarrow}}c^{\dagger}_{
j\downarrow} \ket{\mbox{vac}}, 
\end{eqnarray}
is a member of a spin multiplet of $2 s +1$ states,
$\ket{s,m}^{N}_{0} \propto (S_{-})^{(s-m)}\ket{s,s}^{N}_{0}$,
Here $n_{\uparrow/\downarrow} = N/2 \pm s$, and
$S_{-}= S_x - i S_y$ is the spin-lowering operator.
For $K^{ab}\not=0$ (but still $k^{ab}_{ij} = 0$) the true low-energy
eigenstates, $\ket{s,m}^N$, are linear superpositions of the bare states
in the multiplet $\ket{s,m}^N_0$ (with $\ket{s,m}^N
\to \ket{s,m}^N_0$ as $|K^{ab}|/h \to 0^+$). We shall call the states
$\ket{s,m}^N$ the \emph{spin wave multiplet}, since each can be viewed
as a homogeneous spin wave.  By creating additional single-particle
excitations, other, higher-energy multiplets can be built.
However, their eigenenergies lie higher than those of the spin wave
multiplet $ \ket{s,m}^N $ by an amount
which is at least of order the single-electron level spacing,  i.e.\
 rather large compared to $\bar d_{\rm obs}$ [cf.\ (P1)]; thus
the mechanism causing the observed abundance of low-energy
excitations, whatever it is, must have its origin in spin excitations,
not purely in single-particle excitations.

\begin{figure}
\begin{center}
      {\includegraphics[clip,width=8.6cm]{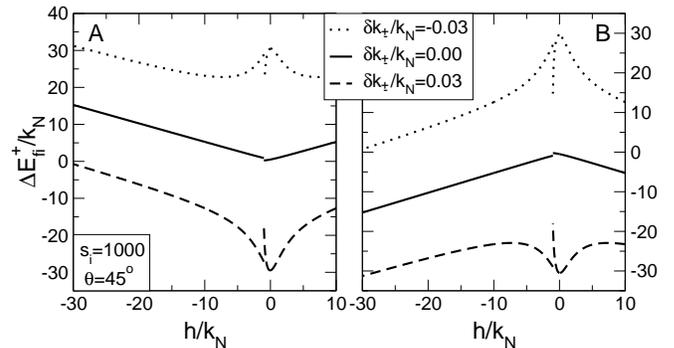}} 
\caption{Tunneling energies $\Delta E_{fi}^{+}$ for 
  ${\cal H}_{\rm Zee} + {\cal H}_{\rm uni}$, plotted as functions of $h/k_N$
  sweeping positive to negative, illustrating the effects of anisotropy
  fluctuations ($\delta k_\pm=k_{N\pm1}-k_N$) for the transitions from
  $\ket{s_i,s_i}^N$ to (A) $ \ket{s_i-{1 \over 2},s_i-{1 \over 2}}^{N+1}$ and
  (B) $\ket{s_i+{1 \over 2},s_i+{1 \over 2}}^{N+1}$. }
\label{fig:spectrum1}
\end{center}
\end{figure}

{\em Anisotropy fluctuations.}--- Let us first turn to the behavior of
the tunneling resonances for small magnetic fields [see (P2)].  The
jumps at the switching field have been attributed to a
sudden reversal of the nanoparticle's magnetic moment \cite{gueron},
which occurs when the energy barrier between a metastable state and
the true ground state is tuned to zero by the applied field.  To
illustrate how such jumps arise in our model, let us (for
simplicity) take the anisotropy to be uniaxial [Eq.~(\ref{eq:uniaxial})]
and consider the case in which the changing magnetic field only
rotates the total spin moment, without changing its magnitude
\cite{spin-not-changed}. We have numerically diagonalized ${\cal
  H}_{\rm Zee} + {\cal H}_{\rm uni}$ as function of $h/k_N$ for
$s_i=1000$ and $s_f=s_i\pm1/2$ to determine the groundstate energies
and the corresponding tunneling energies $\Delta E_{fi}^{\pm}$
(Fig.~\ref{fig:spectrum1}).  The latter indeed show a jump at $h_{\rm sw}$. 
However, if we neglect anisotropy fluctuations by choosing $k_N=k_{N+1}$
(Fig.~\ref{fig:spectrum1}, solid lines), the $\Delta E_{fi}^{\pm}$ lines 
also have two unsatisfactory features: (i) An upward (downward) jump in
$\Delta E^+_{fi} (h)$ as $|h|$ increases past $|h_{\rm sw}|$ is
\emph{always} followed by a positive (negative) large-$h$ slope,
whereas it is observed experimentally (e.g. Fig.~3(a) in
\cite{gueron}) that \emph{either upward or downward} jumps can occur
for states having a given large-$h$ slope; and (ii), beyond the
switching field, the dependence on $h$ is 
monotonic (close to linear),  in
disagreement with recent data  (P3) \cite{mandar}.  All attempts we made to
explain such behavior by choices of $K^{ab}$ corresponding to more
complicated than uniaxial anisotropies, or by higher order terms such
as $K^{abcd}S_a S_b S_c S_d$ \cite{thiaville}, were unsuccessful.

Now, the very fact that the field-dependence of each resonance
in~\cite{gueron,mandar} differs so strikingly from that of all others
implies that the anisotropy energy fluctuates significantly from
eigenstate to eigenstate, which we associate with $k^{ab}_{ij} \neq 0$
in our model.  Although the contribution of such random fluctuations
to the total energy is non-extensive, their effect on energy {\em
  differences} in which extensive contributions largely cancel, can be
very significant.  A detailed statistical analysis of anisotropy
fluctuations is beyond the scope of this paper.  Instead, we shall
mimic the effects of $k^{ab}_{ij} \neq 0$ by simply using two
different anisotropy constants in ${\cal H}_{\rm uni}$, say $k_N$ and
$k_{N \pm 1} \equiv k_N + \delta k_{\pm}$, for $N$- or $(N \pm
1)$-electron states.  $k_N$ can be estimated from the switching field
$k_N\approx \mu_0\mu_B H_{\rm sw}$ [cf. Fig.\ref{fig:spectrum1}]
yielding $ k_N\approx 0.01$meV\cite{previous-estimate-wrong}.  Now, as
illustrated in Fig.~\ref{fig:spectrum1}, $\delta k_\pm / k_N$ in the
range of a few percent is \emph{sufficient to reverse the sign of the
  energy jumps at $H_{\rm SW}$}.  Note that $\delta k_\pm\not= 0$ also
causes the spectral lines to exhibit rather strong non-monotonic
``kinks'' near $h_{\rm sw}$, whose amplitudes are of order $s_0\delta
k_\pm$.  Qualitatively similar non-monotonicities have indeed been
observed recently~\cite{mandar}, with kink amplitudes on the scale of
a few $0.1$ meV, in rough agreement with $s_0\delta k_\pm$ for $s_0
\simeq 1000$.

Anisotropy fluctuations in the range of a few percent are not
unreasonable in nm-scale devices.  Calculations for transition-metal
clusters show that single spin flips can produce a significant change
in the magnetic anisotropy energy \cite{pastor}, and measurements of
Gd clusters indicate that anisotropy energies can vary significantly
for clusters differing only by a single atom \cite{bloomfield}.

We now turn to the low-energy excitations observed
in~\cite{gueron,mandar} (P1). It is natural to
ask~\cite{gueron,canali} whether these might correspond to spin wave
transitions of the form
$\ket{s_i,s_i}_N\rightarrow\ket{s_f,m_f}_{N\pm1}$ 
for different $m_f$ values. However this does
not seem to be the case, for three reasons: (i) It can be shown that
only two transitions (namely  $|s_0, s_0 \rangle \to |s_0 \pm 1/2, s_0
\pm 1/2 \rangle$) have significant weight~\cite{canali,details}.
Resonances associated with final states $|s_f, m_f \rangle$ that
differ only in $m_f$ would (ii) have a spacing of order
$k_N\left[\approx 0.01\right.$meV], which is significantly {\em
  smaller} than observed, and would (iii) exhibit a {\em systematic}
increase in the magnitude of their slope ($\propto | s_i-m_f|$) for
high magnetic fields 
 that was not observed in experiment.

\begin{figure}
\begin{center}
{\includegraphics[width=8.6cm]{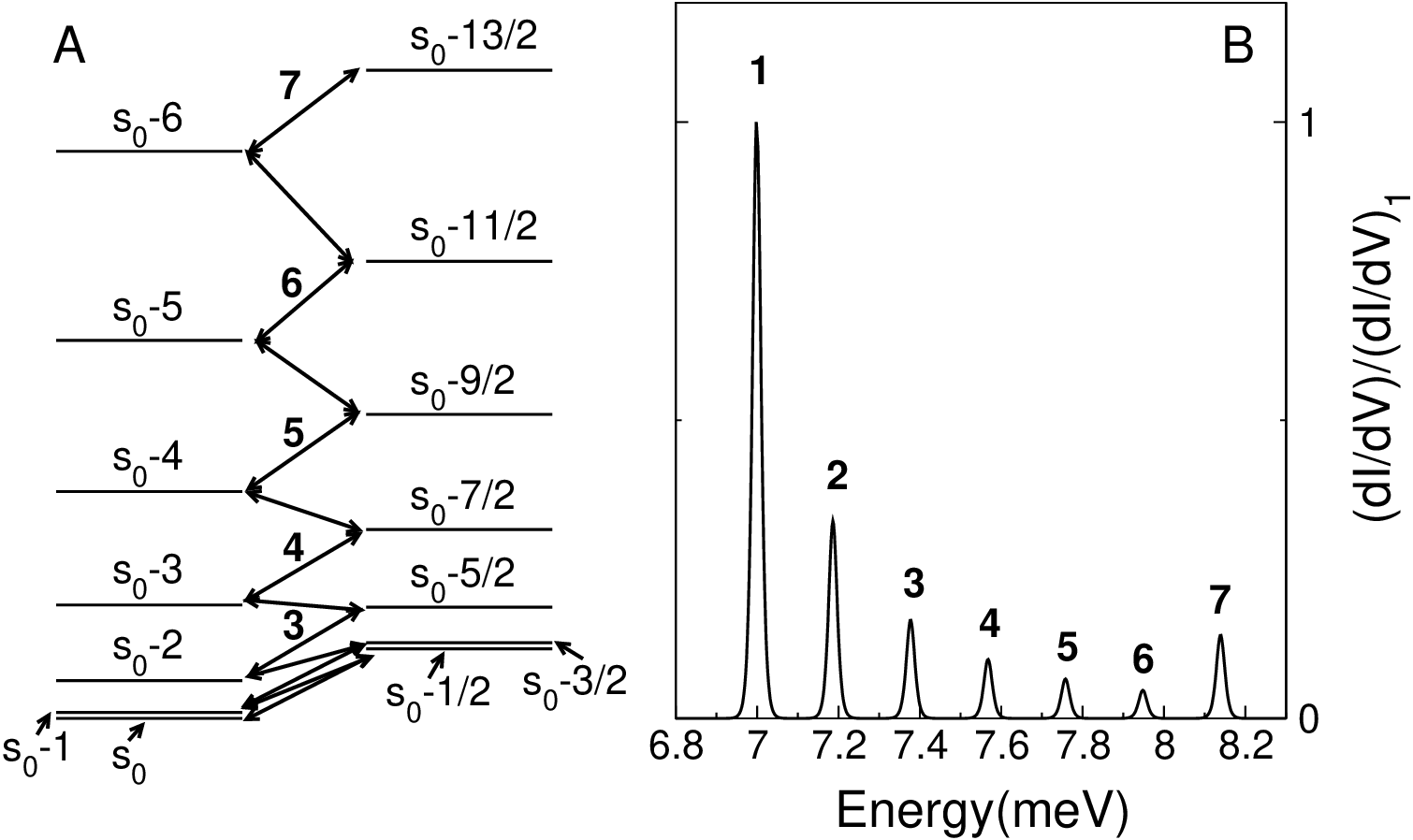}}
\caption{ Nonequilibrium
  spin accumulation in a ferromagnetic nanoparticle: (A) Tunneling transitions
  can cause an energy ladder of states with different total spins to be
  populated ($s_0$ denotes the ground state of the spin-$s_0$ multiplet,
  etc.). (B) The corresponding differential conductance as function of energy,
  normalized by its first maximum and calculated by standard methods
  \protect\cite{barnas-martinek}.  (Parameter choices: $h=0$, ${\cal H}_{\rm
    anis} = 0$, $s_0=1000$, $\Delta E_{0,\rm tot}^{\alpha p}=0$, $T=80$mK,
  $B_p=0.4$, a tunnel junction resistance ratio of $R_L/R_R = 0.1$,
  and a Coulomb blockade threshold of $7$~meV 
(see sample 3 in Fig.~2 in 
\protect\cite{gueron}); we neglected energy and spin relaxation,
and assumed that charging transitions add minority electrons, so that
$\delta E_{\rm res} =\bar d_{\rm min} -U/2$.) 
No significance
should be attached to peak heights here, since they
depend on (unknown) tunneling matrix elements, which
for simplicity we took all equal.
}
\label{fig:ladders}
\end{center}
\end{figure}

{\em Nonequilibrium.}---Since the large density of resonances (P1)
cannot be explained by equilibrium transitions (neither single
particle excitations nor spin wave excitations), we must explore
nonequilibrium effects: In general, $N$-electron states other than the
ground state can be populated during the process of current flow, and
this may affect the experimental tunneling spectrum
\cite{agam,jvdralph}.  Figure~\ref{fig:ladders} illustrates the
consequences as applied to a ferromagnetic grain.  Even if a first
tunneling event causes a ``charging'' transition from the $N$-electron
ground state $| {\rm G} \rangle^N$ to the $(N \pm 1)$-electron ground
state $|{\rm G} \rangle^{N \pm 1}$, it may be energetically possible
for the subsequent ``discharging'' tunneling transition to return the
particle to an excited $N$-electron state $|\alpha \rangle^N$ instead
of $|{\rm G} \rangle^N$, provided the applied voltage is sufficiently
large, $eV \ge E_\alpha^N - E_{\rm G}^N$.  Likewise, further charging
and discharging transitions may allow any of a large ensemble of
states to be occupied at higher and higher levels of an energy ladder,
terminating only when an energy-increasing transition requires more
energy than the applied voltage provides.  As the voltage is
increased, the total current (conductance) may increase stepwise (show
peaks) when thresholds are crossed to allow higher-energy transitions
up the nonequilibrium ladder, thereby changing the occupation
probabilities of the ensemble of nonequilibrium states and opening new
tunneling channels. 

In a
ferromagnetic particle, in addition to the nonequilibrium occupation of
single-electron states discussed previously for nonmagnetic particles
\cite{agam}, nonequilibrium spin excitations are possible, too,
if  the spin-flip rate $\Gamma_{\rm sf}$
   is smaller than the tunneling rate $\Gamma_{\rm tun}$ \cite{rates}.
Fig.\ref{fig:ladders}A illustrates a ladder of transitions between 
ground states $|s,s\rangle$ \cite{quasiparticle}
of multiplets with different total spin $s$, leading 
to {\em spin accumulation} on the grain. Fig~\ref{fig:ladders}B shows the 
corresponding differential conductance, calculated by solving a 
master equation for the population of the states of the ladder.
The resonance peak spacing ($\delta E_{\rm res}$) 
and the number of peaks ($n_{\rm res}$) for such a ladder
can readily be calculated (using Eq.(2) of~\cite{canali}),
and depend on whether the charging transition
adds/removes an electron to/from the grain (to 
be distinguished by an index $p = \pm 1$),
and on whether it is a majority/minority electron (to be distinguished 
by an index $\alpha = \pm 1$). One finds 
$\delta E_{\rm res} = \bar d_{\rm min} -U/2$ for $p \alpha = -1$,
and  $\bar d_{\rm maj} - U/2$ for $p \alpha = 1$. 
Using the model parameters estimated above with $s_0 \approx 1000$,
the first quantity gives a spacing of $\approx 0.2$~meV, as is
observed.  The second quantity is larger, $\approx$
3.6~meV~\cite{spinwaves}.  A detailed analysis~\cite{details} shows
that $n_{\rm res} -1 $ equals the \mbox{smallest integer larger or
  equal to}
\begin{equation}
\label{eq:n_max}
\frac{2E_C^{\rm thresh}-2 B_p\Delta 
E_{0,{\rm tot}}^{\alpha p}}{(B_p-B_{-p})(\bar d_{\rm maj}+
\bar d_{\rm min}-U)-\alpha(\bar d_{\rm maj}-\bar d_{\rm min})}\; , 
\end{equation}
where $B_p=[1+C_p/C_{-p}]^{-1}$ contains the ratio of junction
  capacitances involved in processes $p$ and $-p$, $\Delta E_{0,{\rm
    tot}}^{\alpha p}$ is the energy difference between states $\ket{s_0+\alpha
  p,s_0+\alpha p}$ and $\ket{s_0,s_0}$, and $E_C^{\rm thresh}$ is the
threshold charging energy (energy of the first peak in the differential
conductance).  The prediction that $n_{\rm res}$ increases linearly with
$E_C^{\rm thresh}$~\cite{quasiparticle} is in qualitative agreement with
Fig.~2 of ~\cite{gueron}, and could be checked explicitly in future devices
with gate electrodes, which would allow $E_C^{\rm thresh}$ to be tuned.

A nonequilibrium scenario can also account naturally for the fact (P3)
that the vast majority of the observed transitions within a given
sample shift in energy with a similar slope for large magnetic fields.
This will happen when all the nonequilibrium threshold transitions
correspond to tunneling events with the same change of $S_z$ (see
Fig.~\ref{fig:ladders}), and therefore the same Zeeman shift.

In conclusion, we have proposed a phenomenological model for nanoscale
magnets that treats magnetic interactions within a many-electron
picture.  Its parameters were estimated from bulk properties of Co or
experiment, except for the total spin $s_0$ and the strength of
anisotropy fluctuations, which were used as free parameters.  The
model offers a framework for understanding recent experiments
measuring the discrete excitations of magnetic nanograins, provided
that we assume (i) anisotropy fluctuations of a few percent between
different eigenstates within the same nanograin, and (ii)
nonequilibrium spin accumulation.

We thank C. Canali and A. MacDonald for advance communication of their work,
which significantly influenced parts of ours. Moreover, we acknowledge helpful
discussions with J.  Becker, D.  Boese, E. Bonet Orozco, A. Brataas, E.
Chudnovsky, A.  Garg,  S. Gu\'eron, C. Henley, T. Korb, D.  Loss,  E.
  Myers, W.  M\"orke, A. Pasupathy, J. Petta, and G. Sch\"on.  This
work was supported in part by the DFG through SFB195, the DFG-Pro\-gram
``Semiconductor and Metallic Clusters", the DAAD-NSF, by
US-NSF (DMR-0071631) and the Packard Foundation.

\phantom{.} \vspace*{-1.cm}

\phantom{.} \vspace*{-1cm}

\widetext

\begin{thebibliography}{10}
\bibitem{bulkN} We use $\bar d_{\sigma} = 1/[ N_{\rm a} {\cal N}_\sigma
  (\varepsilon_{\rm F})]$, where ${\cal N}_\sigma (\varepsilon)$ is the
  spin-dependent bulk density of states per atom \cite{Papa}.  Using bulk
  values is appropriate for nanoscale grains, since quantum chemical
  calculations~\cite{rosch} on Ni clusters have shown that ${\cal N}_\sigma
  (\varepsilon)$ has converged to its bulk shape already for $N_{\rm a}$ as
  small as 6.
\bibitem{spin-not-changed} To change the ground state spin value by
  applying a magnetic field requires in general a field of $\delta h
  \sim ( \bar d_{\rm maj}+ \bar d_{\rm min}-U)/2$, which is
  generically (but not necessarily) large in nanomagnets ($\mu_0
  \delta H \gtrsim 25$~T for $s_0 \lesssim   1250$).
\bibitem{previous-estimate-wrong} The estimate $k_N \approx 0.1$~meV
  of Ref.~\cite{gueron} is unreliable, since derived from spectral
  jump sizes, which depend strongly on $k_N$-fluctuations (cf.\ 
  Fig.~\ref{fig:spectrum1}) that were neglected in \cite{gueron}.
\bibitem{rates}  For GDMR's Co nanograins, we estimate $\Gamma_{\rm
    tun}$ from the conductance to be $\Gamma_{\rm tun}\approx G \bar
  d_{\rm min} / e^2 \approx 10^9 {\rm s}^{-1}$, and the rate for
  spin-conserving electronic relaxation (by Eq.(8)  of~\cite{agam})
  to be $\Gamma_{\rm el}\approx 10^9 {\rm s}^{-1}$. Since in general
$\Gamma_{\rm sf} \ll \Gamma_{\rm el}$, we expect
that $\Gamma_{\rm sf} \ll \Gamma_{\rm tun}$ for GDMR's grains.
\bibitem{quasiparticle} For simplicity, we have not included the
      possibility of a nonequilibrium population of single particle
      excitations in our calculations or in Fig.\ref{fig:ladders}.
      However, they certainly exist, and if taken into account, would
      lead to additional resonances in the tunneling spectrum.
\bibitem{spinwaves}  
  Of course, spin wave excitations can be excited too in
  nonequilibrium. However the threshold spacing expected for
  transitions involving spin-wave excitations is $k_N/s_0$ (or
  $2|\delta k_{\pm}|$) if $\delta k_{\pm}=0$ (or $\not=0$) which is
  exceedingly small in both cases.
\end{thebibliography}
\end{document}